\documentclass{PoS}

\title{Status of direct and indirect dark matter searches}

\ShortTitle{Dark matter searches}

\author{\speaker{Carlos P\'erez de los Heros}\\
        Uppsala University\\
        E-mail: \email{cph@physics.uu.se}}

\abstract{I review the current status of dark matter searches using direct, indirect and accelerator techniques.
  A detailed review of individual experiments is beyond the scope of these proceedings. I focus instead on the challenges
  (sometimes limitations) faced by each of the approaches, which is what make them complementary, and the reason we must
 ensure that they are developed concurrently.}

\FullConference{
European Physical Society Conference on High Energy Physics - EPS-HEP2019 -\\
			10-17 July, 2019\\
			Ghent, Belgium}

\begin{document}

\section{Introduction}
The history of how dark matter was ``discovered'' and how the term Dark Matter itself was coined has been told somewhere else~\cite{Sanders:2010cle,Bertone:2016nfn},
and it is beyond the scope of these proceedings to delve further on it. There is, however, a turning point somewhere between the 1930's, when
the first indications of an anomaly in the rotation curves of galaxies was assigned to ``extinguished stars, dark clouds, meteors, comets and so on''~\cite{Lundmark:30aa},
and the present day, where we know that dark matter can not be made up of normal, that is baryonic, matter.

Several microlensing experiments during the 90s and 2000s~\cite{Alcock:2000ph,Tisserand:2006zx,Wyrzykowski:2011tr} found that the amount of non-luminous objects in the halo of the Milky Way
lies at the few percent level,
insufficient to account for the additional matter needed to explain the rotational curve of the galaxy. Around the same time, precise estimations of the
amount of primordial light elements from big-bang nucleosynthesis set a very precise limit on the total amount of baryons in the Universe (see~\cite{Cyburt:2015mya} for a review). 
Further, the first measurements of the Cosmic Microwave Background (CMB) radiation from space with the COBE mission in the 90s~\cite{Smoot:1998jt}, improved later with
WMAP~\cite{Hinshaw:2012aka} and Planck~\cite{Aghanim:2018eyx}, showed the need for a non-baryonic component in the matter budget of the universe.

The problem then became, and remains, to identify and detect this new form of matter. The easiest solution, at least from the point of view of a particle physicist, is
to introduce a new particle species that needs to be heavy (we want it to play a role in gravitational structure formation), 
weakly interacting (we do not want it to disturb the evolution of the Universe) and stable (its effects must perdure today). Other solutions avoiding dark matter, like modifying the
classical gravitational interaction for low accelerations, Modified Newtonian Dynamics (MOND)~\cite{Famaey:2011kh}, or its relativistic extensions, although able to explain remarkably well the rotation
curves of galaxies, are faced with difficulties when trying to become global explanations, that is, to explain the dark matter distribution at the galaxy-cluster level.
We will not describe these approaches here, but concentrate on the standard weakly--interacting new particle paradigm.

\begin{figure}[t]
  \begin{minipage}{0.47\linewidth}
		\includegraphics[width=\textwidth]{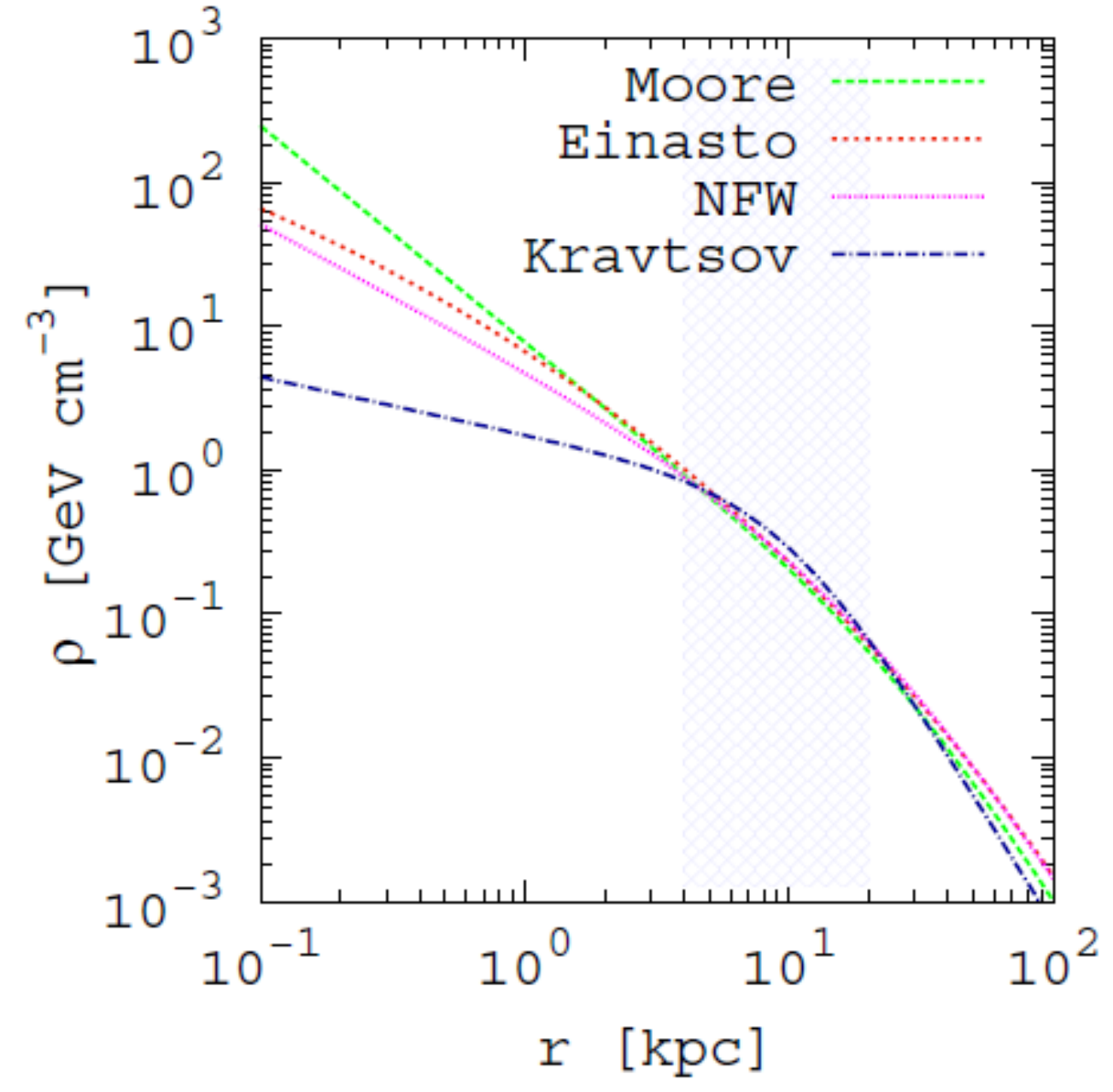}
	\caption{The density of dark matter as a function of distance to the galactic center in commonly assumed dark matter halo density profiles for the Milky Way. Reprinted with permission from~\cite{Abbasi:2011eq}.}
	\label{fig:profiles}
  \end{minipage}
  \hfill
  \begin{minipage}{0.47\linewidth}
    \vspace{-0.4cm}
    \includegraphics[width=\textwidth]{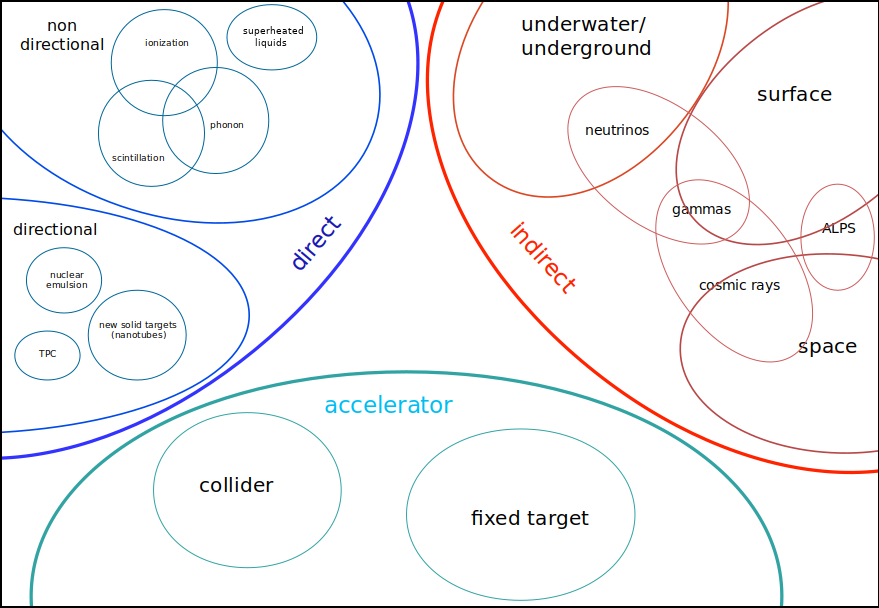}
    \vspace{1.cm}
	\caption{Current approaches to dark matter detection.}
	\label{fig:techniques}
\end{minipage}
\end{figure}

Any dark matter candidate particle, $\chi$, will be in thermal equilibrium in the early universe as long as the reaction $\chi\chi \rightarrow SM SM$ holds, SM being any relevant
Standard Model particle that couples to the $\chi$. As the universe expands, the equilibrium will last until the $\chi$'s are diluted so that the above reaction becomes unlikely, leaving
a relic density of $\chi$'s, whose value depends on the cross section that drives the equilibrium condition. The relic density of $\chi$'s, $\Omega_{\chi}$, as a function of their annihilation
cross section, $\sigma_{\chi\chi}$, can be expressed as $\Omega_{\chi} h^2 \propto {10^{-26}(\rm{cm^3/s)}}/{\left < \sigma_{\chi\chi} v \right >}$~\cite{Garrett:2010hd}.
The strength of the needed cross section is of the order of the weak interaction, so the particle solution to the dark
matter problem provides then an
intriguing link between the Standard Model of Particle Physics (SM) and the $\Lambda$CDM model, the standard model of cosmology.  $\Lambda$CDM needs the known elementary particles and forces
from the SM, a weakly-interacting, stable, cold dark matter candidate (or candidates) and a cosmological constant. With these ingredients, numerical simulations exactly predict the grow of
fluctuations from the early universe into a large-scale structure of galaxies compatible with observations~\cite{Planelles:2014zaa}, a triumph of $\Lambda$CDM. A key point worth mentioning
is that the same numerical simulations but without a dark-matter component, fail to reproduce the large-scale universe as we know it, which can be taken as additional ``evidence'' for dark matter. 
Similarly, N-body simulations of galaxy formation tell us that galaxies are embedded in clumpy, complex dark matter halos that extend well beyond the visible galaxy. Understanding halos is key
to interpreting experimental results, as we will mention below. There have been several proposals to describe the dark matter density distribution in galactic halos as a function of the distance from the center of the galaxy~\cite{Navarro:1995iw,Einasto:1965czb,Moore:1999gc,Kravtsov:1997dp}, but the different models can be parameterized with a single generic function~\cite{Zhao:1995cp},
\begin{equation}
\rho_{DM}(r)\,=\,\frac{\rho_0}{\left ( \delta + \frac{r}{r_s} \right )^\gamma \cdot \left ( 1 + (\frac{r}{r_s})^\alpha   \right )^{(\beta-\gamma)/\alpha} }
\label{eq:profiles}
\end{equation}
where $r$ is the distance from the galactic center, $\rho_0$ is a normalization constant, and the parameters $\alpha$, $\beta$ and $\gamma$ determine the shape of the halo. Different combinations
of these parameters can recover the standard halo profiles proposed in~\cite{Navarro:1995iw,Einasto:1965czb,Moore:1999gc,Kravtsov:1997dp}, and shown in Figure~\ref{fig:profiles}. But equation~(\ref{eq:profiles}) can easily incorporate new parametrizations. These profiles describe the smooth distribution of dark matter around galaxies. A possible clumpy component must be added on top in order to describe the outcome of N-body simulations.

Figure~\ref{fig:profiles} shows that there is consensus on the dark mater distribution at the location of the Solar System (about 8 kpc from the galactic centre), but
predictions diverge considerably at distances near the center of the galaxy. This is of course also true when considering other galaxies and assuming the same
type of profiles apply universally.

\section{Dark matter detection techniques and results}
Figure~\ref{fig:techniques} shows schematically the different experimental techniques currently used, or under development, to detect dark matter. We can distinguish
three generic complementary approaches: direct detection, indirect detection and production in the laboratory.
Within each group one can identify a wealth of different detection methods, targets or type of signals. 
To describe each of them lies outside the scope of these proceedings. In what follows we will just touch upon a few distinct aspects of each group and show the
current experimental status.

\subsection{Direct searches}
Direct searches are based on looking for nuclear recoils from dark matter-nucleus interactions in a suitable target. The expected number of recoils (i.e., interactions)
of a dark matter particle of mass $m_{\chi}$ on a target made of particles of mass $m_A$ is given by~\cite{Schumann:2019eaa}
\begin{equation}
 \frac{dR}{dE}(E,t)\,=\,\frac{\rho_0}{m_{\chi} m_A}\, \int v\cdot f(v) \frac{d\sigma}{dE}(E,v) d^3 v,
\label{eq:recoils}
\end{equation}
where $\rho_o$ is the local dark matter density, $f(v)$ is the dark matter velocity distribution and $\sigma$ is the
dark matter-nucleus cross section.

Equation~(\ref{eq:recoils}) has too many unknowns to be useful out of the box: $\rho_o$,  $f(v)$,
$m_{\chi}$ and $\sigma$. So a model for the astrophysics input, $\rho_o$ and $f(v)$, needs to be assumed, and results are then expressed in
terms of $m_{\chi}$ versus $\sigma$. The choice of target ($m_A$) is important because, along with the location of the detector
to reduce background, is the only handle an experimentalist has on the above equation. The choice of target influences also which kind of 
interactions an experiment is more sensitive to.  The cross section can be divided into a spin-dependent component, $\sigma^{\mathrm{SD}}_{\chi\mathrm{-}p}$,
reflecting the coupling from axial-vector terms of the Lagrangian, and a spin-independent component, $\sigma^{\mathrm{SI}}_{\chi\mathrm{-}p}$, from scalar
and vector couplings in the Lagrangian. Targets with large angular momentum provide sensitivity to the spin-dependent part, while larger targets
favour the spin-independent part, which is just proportional to the number of nucleons in the system. Note also that both the quark content of
the nucleon and the nucleon distribution of target nuclei play an essential role in calculating observables and interpreting experimental results, and they can be
a source of uncertainty in the quoted limits or in comparisons with other experiments~\cite{Bottino:1999ei,Bottino:2001dj,Ellis:2008hf,deAustri:2013saa,Hoferichter:2018acd}.

The background in a nuclear recoil experiment arises from
radioactivity in the surroundings or from the materials of the detector itself. The experimental efforts in the last couple of decades have
been focused in achieving extremely stable, radio-pure, shielded detectors, with an energy threshold as low as possible (typically a few keV). 
As detectors increase sensitivity to lower recoil energies, the background from elastic neutrino scattering on target electrons, or coherent neutrino
scattering on target nuclei becomes an issue. This is an irreducible background since even underground sites in deep mines are subject to a continuous
flux of atmospheric and solar neutrinos. Usually called the ``neutrino floor'', this background can be dealt with by developing detectors sensitive to the direction
of the recoil, and therefore to the direction of the incomming dark matter particle. This is a relatively new approach and there is a wealth of R\&D in this 
direction~\cite{Mayet:2016zxu}. Directional sensitivity will give a handle to reject events from the direction of the Sun, 
most likely induced by a solar neutrino, and will give
the possibility to better exploit the expected annual variations in the recoil rate due to the relative velocity of the Earth in the 
dark matter halo. Depending on the location of the detector, a daily modulation in the recoil rate (day-night effect), induced by the variation of the
relative velocity with the dark matter as the Earth rotates, can also be expected. \\

\begin{figure}[t]
  \begin{minipage}{0.47\linewidth}
		\includegraphics[width=\textwidth]{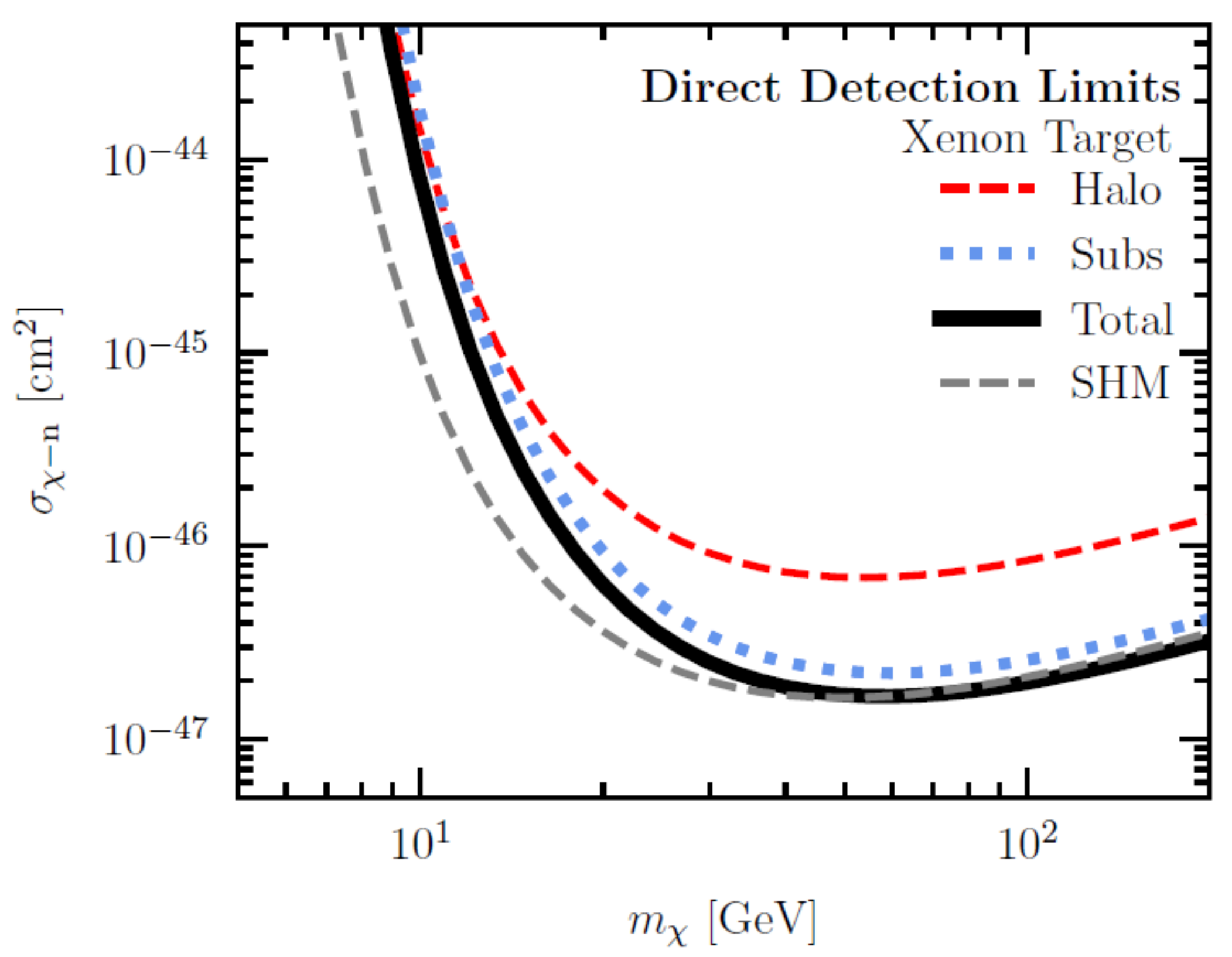}
	\caption{95\% background-free C.L. limits on the DM-nucleon spin-independent scattering cross section as a
        function of DM mass. A background-free, Xenon target experiment with an exposure of 1 kton x year and a 4.9 keV energy threshold for the nuclear recoil was
          assumed as benchmark. ``SHM'' stands for Standard Halo Model, while ``Total'' assumes the new velocity distribution extracted from Gaia and SDSS data.
          Figure from~\cite{Necib:2018iwb}. Copyright AAS. Reproduced with permission.}
	\label{fig:fv}
  \end{minipage}
  \hfill
  \begin{minipage}{0.47\linewidth}
    \includegraphics[width=\textwidth]{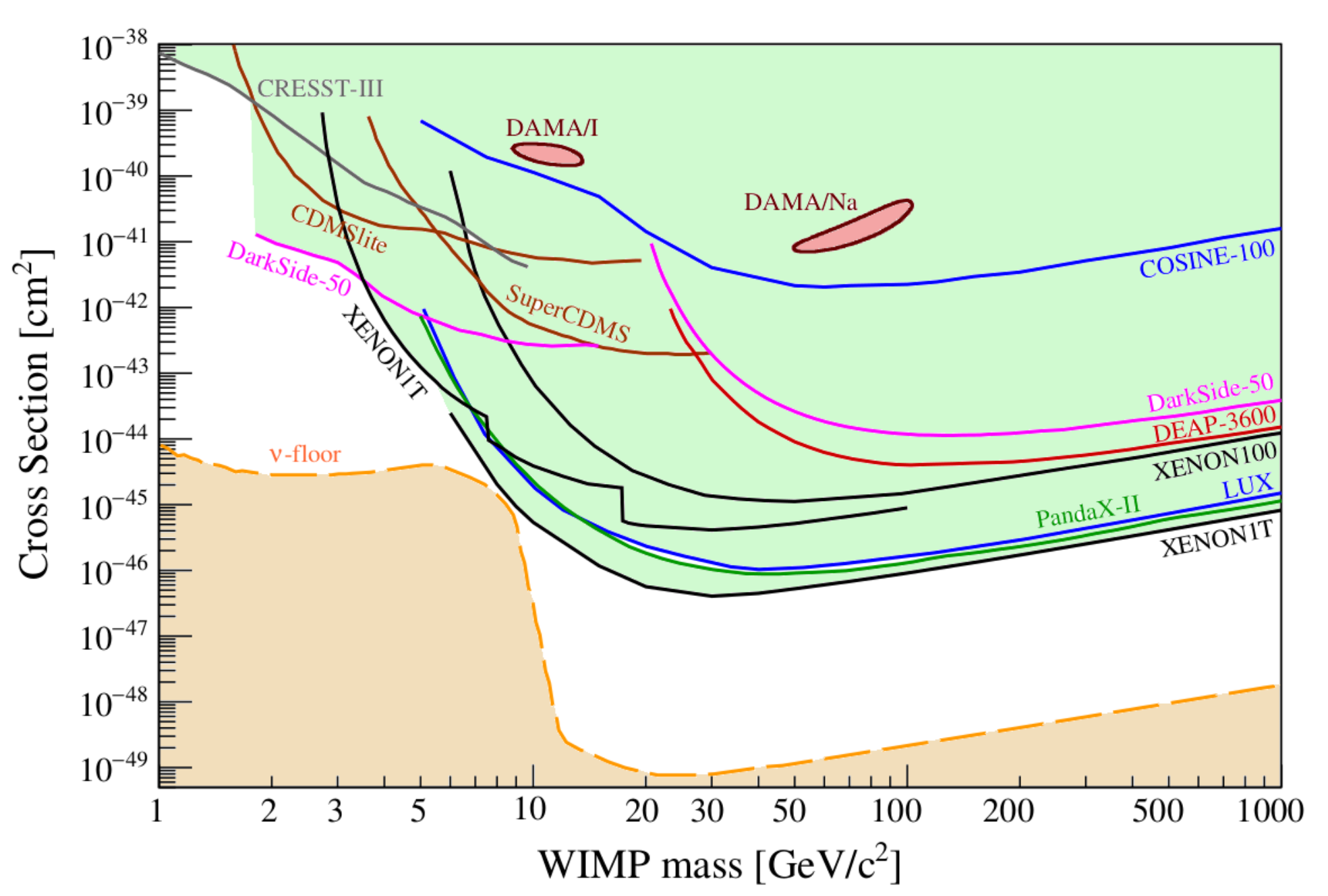}
    \caption{Current experimental limits on spin-independent dark matter-nucleon cross section. Parameter combinations above the lines (i.e., the green shaded area)
      are disfavoured at 90\% confidence level. The dashed line represents the neutrino floor. The regions labeled ``DAMA'' mark the preferred parameter space if 
      the annual modulation seen by DAMA/LIBRA~\cite{Bernabei:2018yyw} would be interpreted as originating from dark matter interactions. Figure courtesy of M. Schumann.
    Updated from~\cite{Schumann:2019eaa}.}
	\label{fig:DD_limits}
\end{minipage}
\end{figure}

 The assumed dark matter velocity distribution can have also other consequences in the interpretation of direct detection results~\cite{Kuhlen:2009vh,Necib:2018iwb,Wu:2019nhd}. 
Note that direct detection experiments are sensitive to the high-velocity tail of the, really unknown, $f(v)$ distribution (high-energy particles produce
stronger recoils in the target, easier to detect). So, different assumptions on $f(v)$ bear on the final result, as shown in Figure~\ref{fig:fv}. The figure
shows expected limits on the DM-nucleon spin-independent scattering cross section assuming a standard Maxwell-Boltzmann velocity distribution in equation~\ref{eq:recoils},
as compared with a distribution derived from recent Gaia and Sloan Digital Sky Survey (SDSS) data (see~\cite{Necib:2018iwb} for details of underlying assumptions on detector
performance). As expected from simple kinematics, suppressing the high energy tail of the assumed $f(v)$ results in a worse sensitivity of the detector to low DM masses. 

Figure~\ref{fig:DD_limits} shows a summary of results from current direct detection experiments. The plot shows the limits on the spin-independent dark matter-nucleon cross section as a function of dark  matter mass. Regions above the curves are disfavoured at  90\% confidence level. The dashed line represents the neutrino floor, the level where neutrino coherent scattering on the detector target nuclei becomes an irreducible background. The regions labeled ``DAMA'' mark the preferred parameter space if the annual modulation seen by DAMA/LIBRA~\cite{Bernabei:2018yyw} would be interpreted as originating from dark matter interactions. These regions are however disfavoured by every single one of the other experiments with sensitivity to that region, so accomodating the DAMA claim that the detected annual modulation is due to dark matter is challenging. It would require extremely ad-hoc interactions between dark matter and baryons in order for the signal to have escaped all other experiments. The measurement remains unexplained, though several ideas have been proposed. 
The plot shows that current experiments quickly loose sensitivity for very low dark matter masses due to threshold effects. Here is where the field faces a technological challenge in the coming years. Using targets and detection techniques that provide sensitivity to electron recoils (in e.g., semiconductors, noble liquids, carbon nanotubes), rather than to nuclei, is a way forward to increase sensitivity to lower dark matter masses~\cite{Crisler:2018gci,Cavoto:2017otc}. Electron recoils have more complex kinematics than nuclear recoils and the detection is challenging. 
Other proposals rely on producing sub-GeV dark matter particles in dedicated beam-dump accelerator experiments, using missing momentum and energy techniques (see section~\ref{sec:colliders} below).  

\subsection{Indirect searches}
Indirect searches for dark matter focus on detecting an anomalous flux of photons, neutrinos or cosmic rays produced in annihilations or decay of dark matter particles gravitationally accumulated in heavy objects, like galaxies, the Sun or the Earth. Detecting the different signatures require very different types of detectors: air shower arrays, Cherenkov telescopes, neutrino telescopes or particle detectors in balloons or satellites. Note that these kind of detectors were not originally intended to search for dark matter but have proven to be unique complementary tools to the direct search efforts. For one thing, they can probe a different side of the velocity distribution of galactic dark matter (low velocity dark matter particles are more likely to be captured in the Galactic center, the Sun or the Earth). For another, they are sensitive to different backgrounds and systematics than direct search experiments. They are also sensitive to the signatures of dark matter decay, unlike direct searches.

There are two sources of possible dark matter signatures where only neutrino telescopes are of use: annihilations in the center of the Sun or Earth. Among the dark matter annihilation products, only neutrinos will escape the dense interiors of these objects. These are also one of the mot background-free searches possible since neither the Sun or the Earth are expected to be sources of high energy (above GeV) neutrinos (except for neutrinos produced in cosmic ray interactions in the atmosphere of the Sun, which can constitute in principle a background to dark mater searches~\cite{Seckel:1991ffa,Moskalenko:1993ke,Ingelman:1996mj,Ng:2017aur,Edsjo:2017kjk}). Assuming equilibrium between capture and annihilation, the neutrino flux from the Sun, $d\Phi_{\nu}/dE_{\nu}$, is proportional to the annihilation rate of dark matter, $\Gamma_A$, which in turn can be related to the capture cross section, that is, the dark matter-nucleon cross section~\cite{Jungman:1995df}.

\begin{equation}
 \frac{d\Phi_{\nu}}{dE_{\nu}}\,=\,\frac{\Gamma_A}{4\pi D^2}\,\frac{dN_{\nu}}{dE_{\nu}}
\label{eq:indirect_sun}
\end{equation}
where $dN_{\nu}/dE_{\nu}$ is the neutrino spectrum from the annihilations.

\begin{figure}[t]
  \begin{minipage}{0.47\linewidth}
   \includegraphics[width=\textwidth]{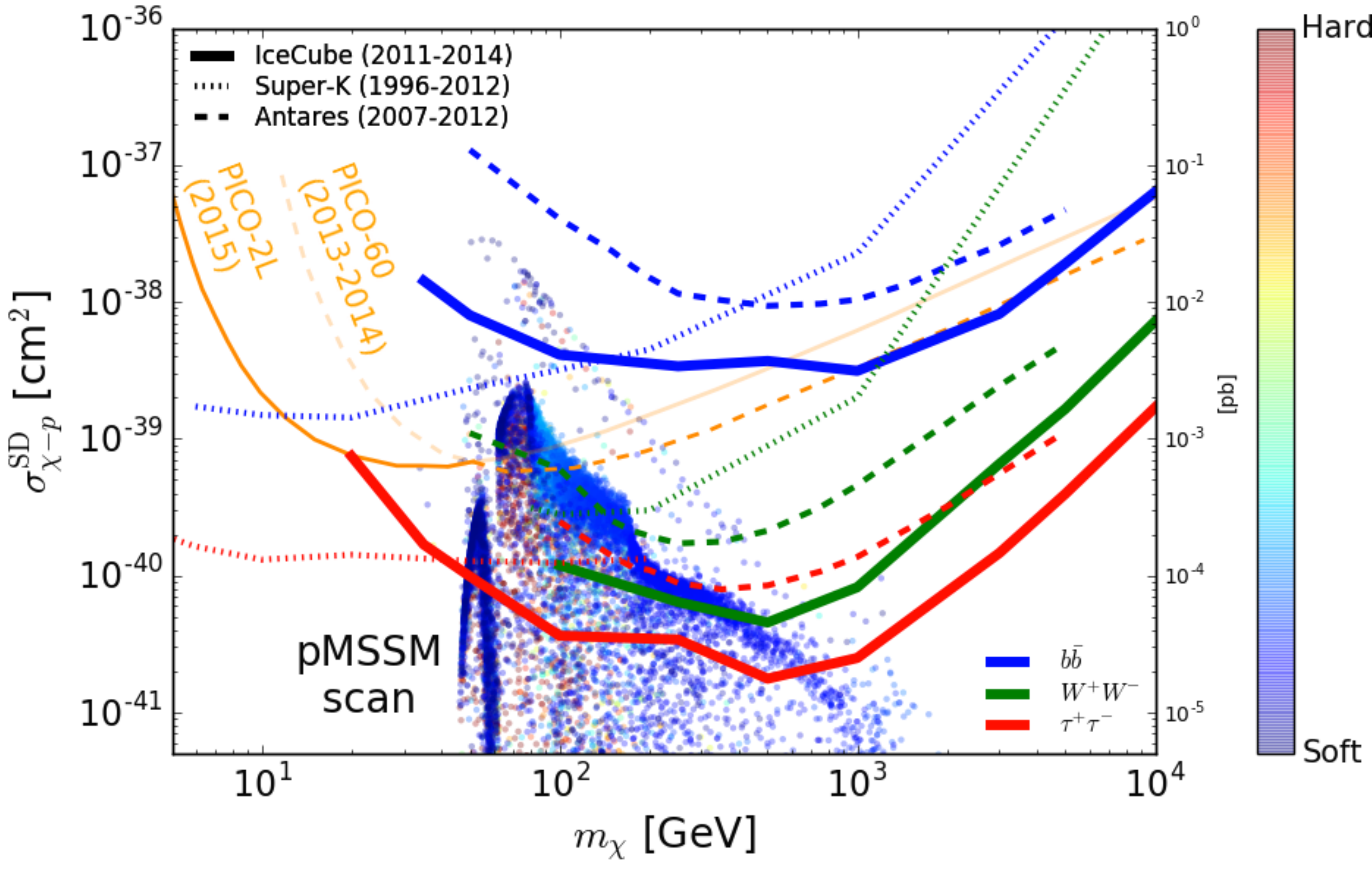}
    \caption{Limits on the spin-dependent dark matter-proton cross-section ($\sigma^{\mathrm{SD}}_{\chi\mathrm{-}p}$), compared to results from other neutrino detectors and direct detection experiments. Various points corresponding to neutralinos from a scan of the phenomenological minimally supersymmetric standard model (pMSSM) are also shown, colour coded by their leading annihilation channel. Points close to the red end of the spectrum annihilate into harder channels such as $\tau^+\tau^-$  and can be excluded by the red line from IceCube. Figure from~\cite{Aartsen:2016zhm}.}
	\label{fig:solar_limits}
  \end{minipage}
  \hfill
  \begin{minipage}{0.47\linewidth}
    \includegraphics[width=\textwidth]{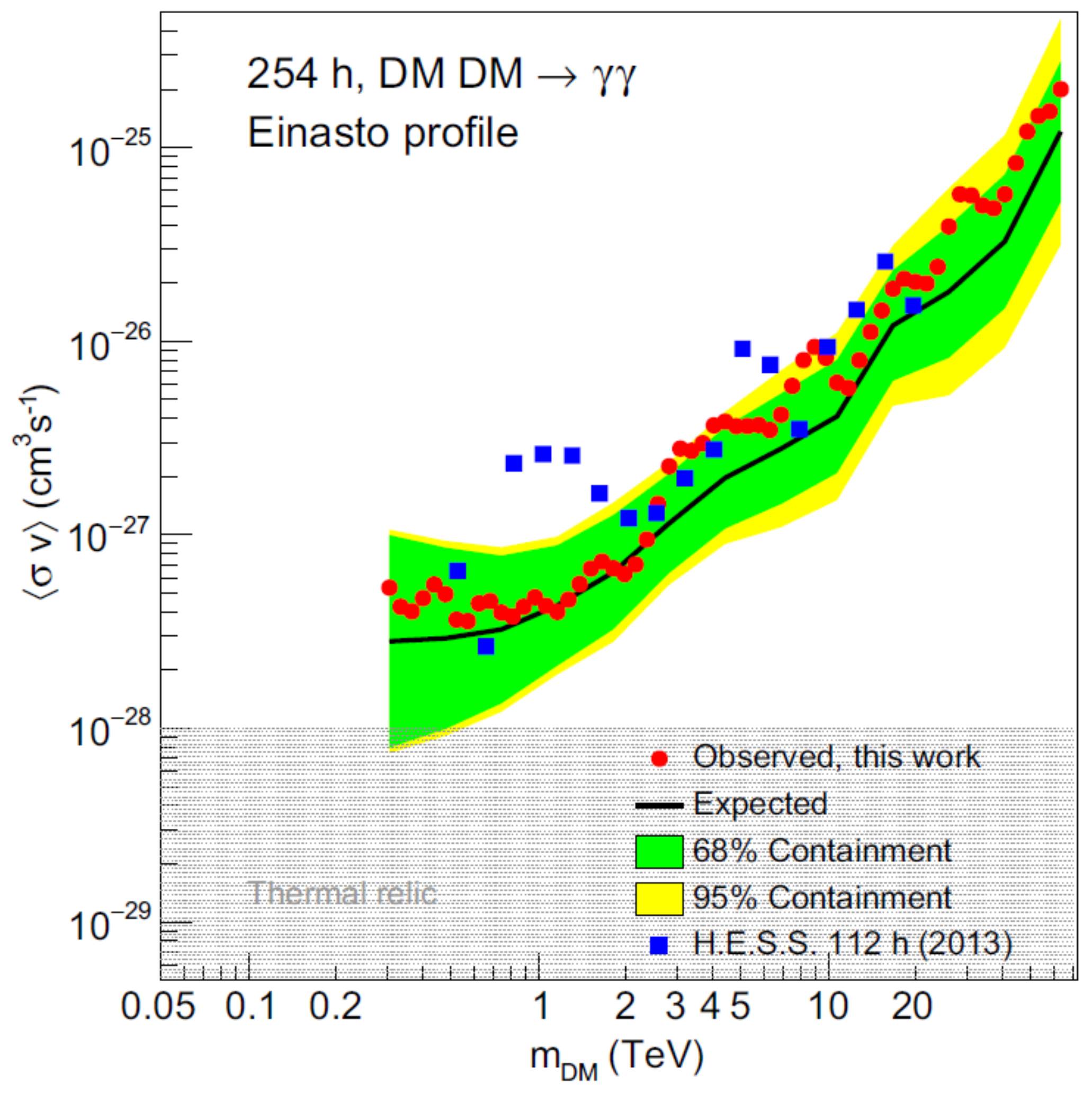}
    \caption{95\% C. L. upper limits on  $\left < \sigma_{\chi\chi} \textrm{v} \right >$ as a function of dark matter mass from the H.E.S.S. experiment, assuming dark matter annihilation into photons. The result is based on observations of the inner 300 pc of the galactic center region. Observed
limits (red dots) and mean expected limit (black solid line) are shown together with the 1$\sigma$ (green band) and 2$\sigma$
(yellow band) containment bands. Figure from~\cite{Rinchiuso:2019rrh}.}
	\label{fig:HESS_limits}
\end{minipage}
\end{figure}

Since the Sun is essentially a target made of protons, it is the spin-dependent cross section that can be measured (or constrained) from a detection (or non-detection) of an anomalous neutrino flux from the Sun.  Figure~\ref{fig:solar_limits} shows the limits obtained by IceCube, Super-K and ANTARES on the spin-dependent dark matter-nucleon cross section as a function of dark matter mass, assuming full annihilation to $b\bar{b}$, $W^+W^-$ and $\tau^+\tau^-$~\cite{Aartsen:2016zhm}. For the Earth, being much younger than the Sun, it is not obvious that equilibrium between capture of surrounding dark matter particles and annihilation in its interior has reached equilibrium. In this case one can still constrain the dark-matter nucleon cross section, but under the assumption of a given value for the annihilation cross section. Since the most abundant isotopes in the Earth inner core, mantle and crust are spin-0 nuclei ($Fe$, $Si$ and $O$), it is the spin-independent dark matter-nucleon cross section that is probed in dark matter searches from the Earth with neutrino telescopes.

Things are a bit different in searches for dark matter from the galactic center or halo, or other galaxies. The neutrino, gamma or cosmic-ray flux from dark matter
annihilations in those cases depends on the thermally averaged product of the dark matter self-annihilation cross-section times the dark matter velocity, $\left < \sigma_{\chi\chi} v \right >$, and the so called $J$-factor, the integral of the squared of the dark matter density along the line of sight to the object under consideration,

\begin{equation}
 \frac{d\Phi_x}{dE_x}\,=\,\frac{1}{4\pi}\,\frac{\left < \sigma_{\chi\chi} v \right >}{2 m^2_{\chi}}\,\frac{dN_x}{dE_x} \times \int_{l.o.s.} \rho^2_{DM}(r) dr d\Omega
\label{eq:indirect_galaxy}
\end{equation}
where $x$ stands for neutrinos, gamma rays or cosmic rays. 
The $J$-factor depends on the halo profile chosen (see equation (\ref{eq:profiles})), and results from indirect dark matter searches are commonly given under the assumption of a specific halo model. Assuming a particle physics model which gives the expected particle spectrum, $dN_x/dE_x$, and a halo model, then an experimental measurement of $d\Phi_x/dE_x$ can be used to probe $\left < \sigma_{\chi\chi} \textrm{v} \right >$ versus the dark matter mass, $m_{\chi}$. An example of a search for dark matter annihilations into photons in the galactic center is shown in Figure~\ref{fig:HESS_limits}, which corresponds to a recent search by the H.E.S.S. collaboration~\cite{Rinchiuso:2019rrh}.

\begin{figure}[t]
  \begin{minipage}{0.47\linewidth}
   \includegraphics[width=\textwidth]{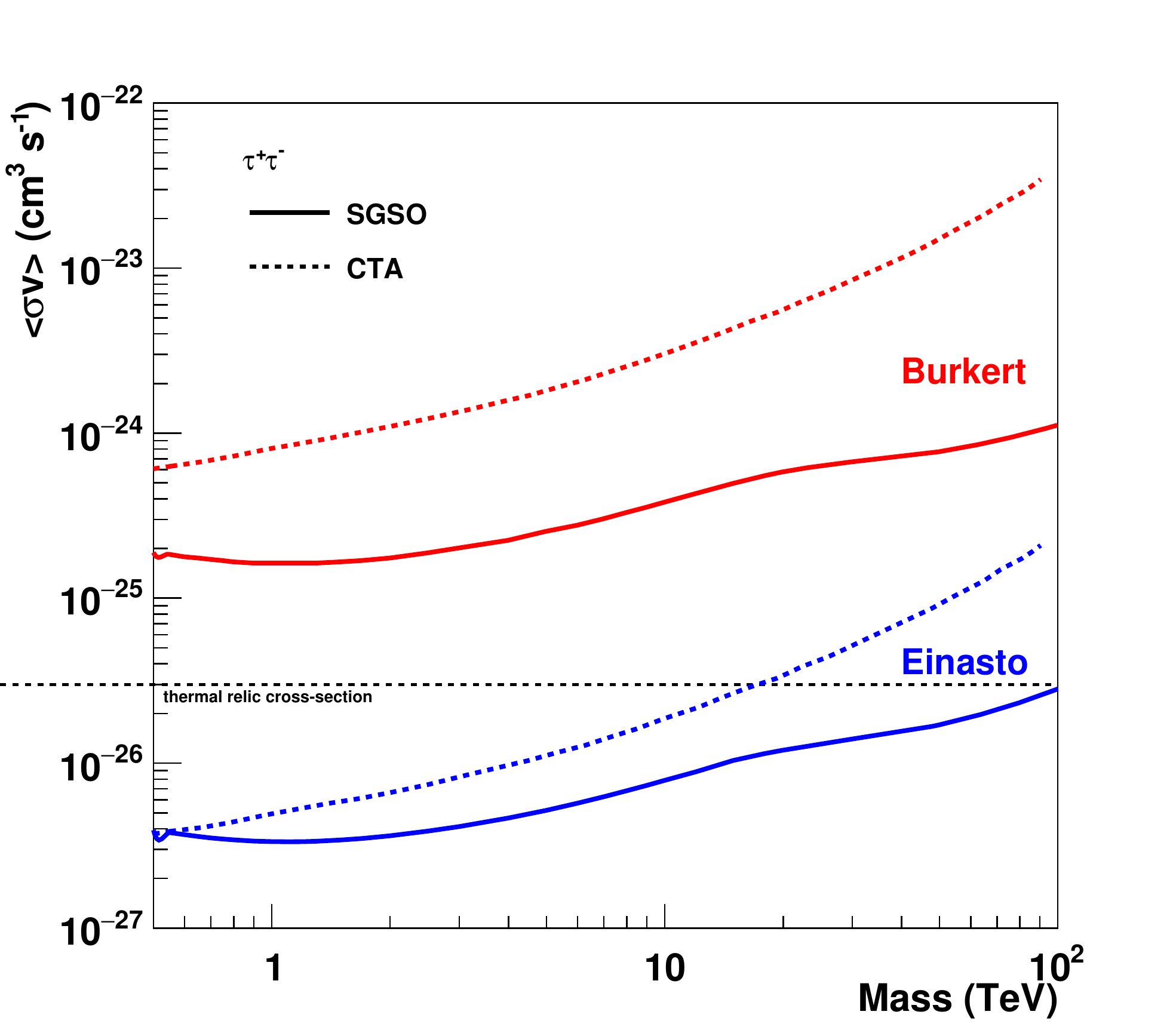}
    \caption{95\% C.L. sensitivity on the thermally averaged cross section for dark matter 
annihilation into $\tau^+\tau^-$ as a function of dark matter mass, for both Einasto  and Burkert profiles of the Galactic halo. See~\cite{Viana:2019ucn} for details.}
	\label{fig:halo_model}
  \end{minipage}
  \hfill
  \begin{minipage}{0.47\linewidth}
    \includegraphics[width=\textwidth]{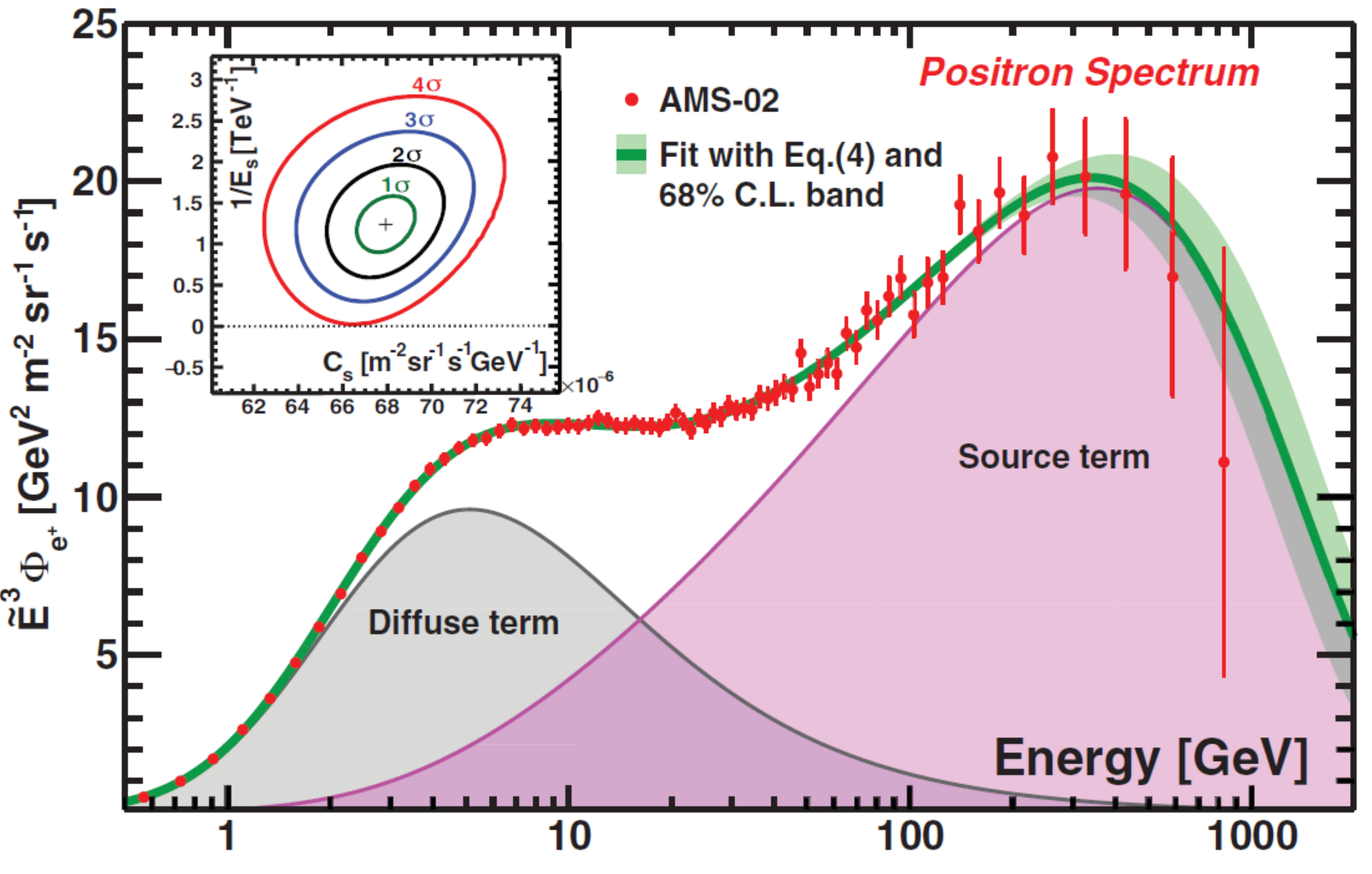}
    \caption{Positron flux as a function of energy measured by AMS~\cite{Aguilar:2019owu}. The red data points show the measured positron flux.
      The data can be fitted with a diffuse low-energy term contribution, explained with positrons produced in the collisions of cosmic rays with
      interstellar gas, and a yet-unidentified contribution at higher energies denoted ``source term'' in the plot.}
	\label{fig:ams_positrons}
\end{minipage}
\end{figure}

The uncertainty introduced by the choice of halo model in indirect dark matter searches is well illustrated in Figure~\ref{fig:halo_model}. The figure shows the sensitivity of two planned facilities, the Cherenkov Telescope Array (CTA)~\cite{CTA:2019aaa} and the Southern Gamma-ray Survey Observatory (SGSO)~\cite{Albert:2019afb} under the assumption of two different halo models. The difference can be orders of magnitude.

Cosmic rays have given so far room for speculation since the detection of a positron excess by PAMELA~\cite{Adriani:2008zr} (with some weak indications from previous experiments) and confirmed by AMS~\cite{Aguilar:2019owu}. The positron spectrum measured by AMS is shown in Figure~\ref{fig:ams_positrons} and can be explained as a low energy part originating from the collisions of cosmic rays in the interstellar medium, and a high energy part of unclear origin. Explanations involving dark matter annihilations have been proposed (the literature is too vast here, but see, e.g.,~\cite{Bergstrom:2008gr,Kopp:2013eka}), but conventional explanations based on positron production by astrophysical sources such as pulsars or supernova remnants are also possible~\cite{Yuan:2013eja,Kohri:2015mga}. Or not~\cite{Abeysekara:2017old}. Use Occam's razor at your discretion. It has to be pointed out that any non-conventional explanation of the positron excess in cosmic rays has to agree with the fact that the antiproton flux seems to be quite well understood (see however~\cite{Lin:2016ezz,Cuoco:2016eej}), making the dark matter explanation of the positron excess quite contrived. We indeed need a complete picture of the antimatter in cosmic rays, including precise measurements of $He$ and $d$ with the spectra extending beyond the TeV region to be able to assess whether there is any cut-off in the spectra.

\subsection{Collider searches}
\label{sec:colliders}
Can we eventually find dark matter in colliders? Strictly speaking: not really. We can find new particles that could do as dark matter, but we would need external
confirmation from Astroparticle experiments to really determine that they are the stuff that holds galaxies together. Still the idea of producing the dark matter
particles in the controlled environment of an accelerator and be able to measure their properties is appealing. 

\begin{figure}[t]
  \begin{minipage}{0.47\linewidth}
   \includegraphics[width=\textwidth]{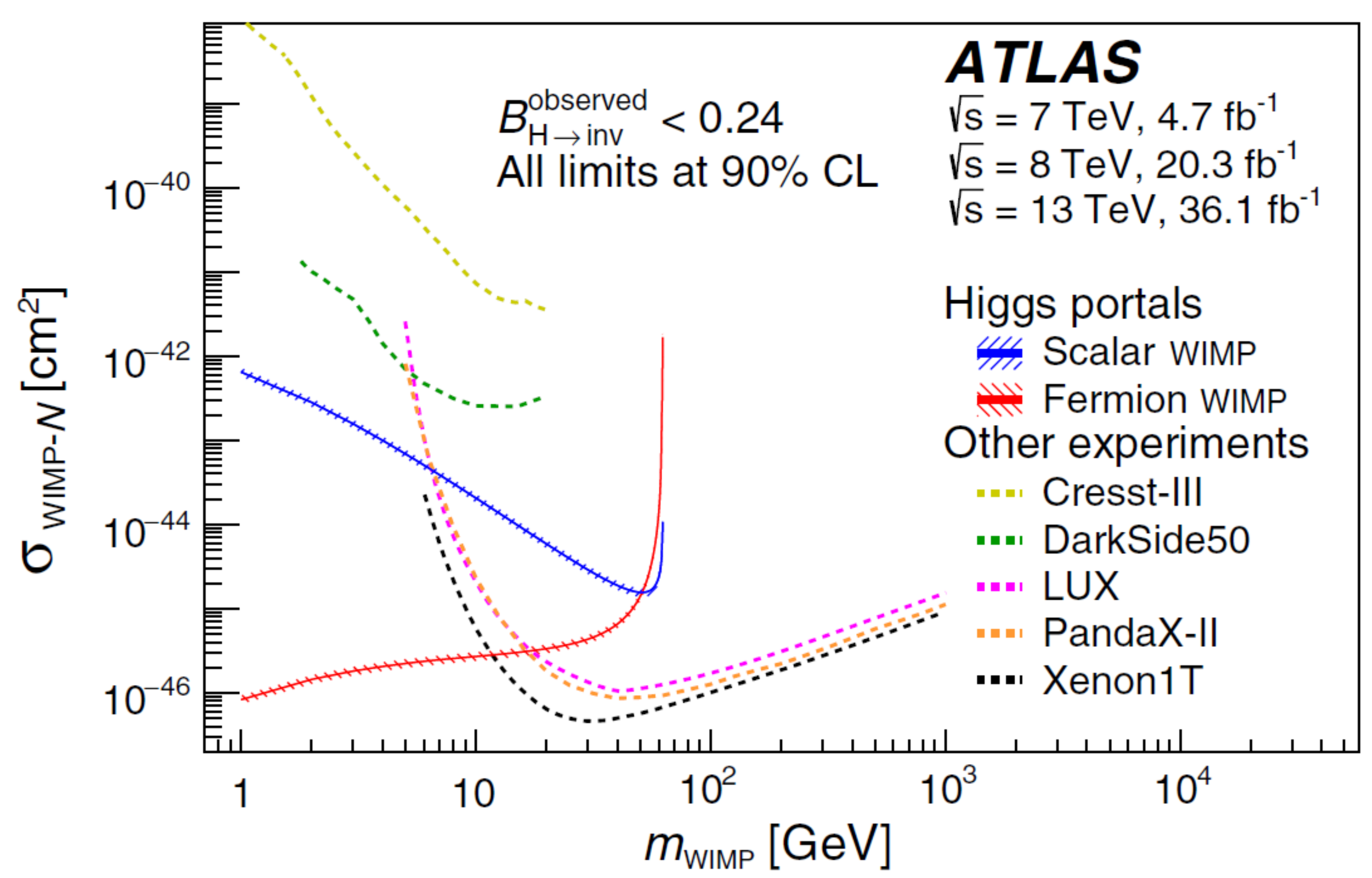}
   \caption{Comparison of the upper limits at 90\% C.L.  on the $\sigma^{\mathrm{SI}}_{\chi\mathrm{-}N}$ obtained by ATLAS, compared with
     several direct detection results. The regions above the contours are excluded. See~\cite{Aaboud:2019rtt} for details.}
	\label{fig:ATLAS_DM}
  \end{minipage}
  \hfill
  \begin{minipage}{0.47\linewidth}
    \includegraphics[width=\textwidth]{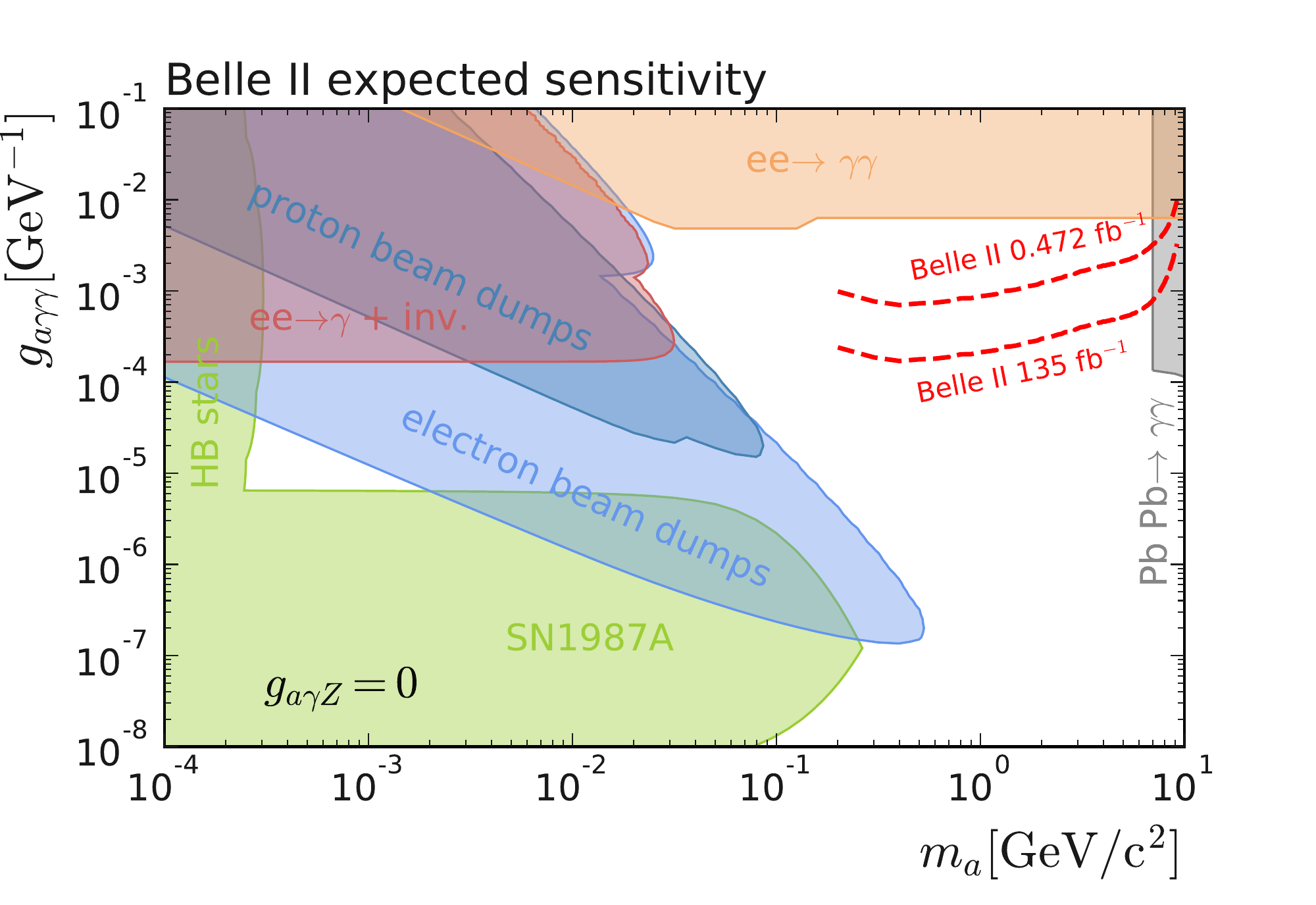}
    \caption{Sensitivity of the Belle-II detector to the axion-photon coupling as a function of the axion mass, compared with disfavoured regions obtained with other technoques. Figure courtesy of S. Cunliffe. Adapted from~\cite{Dolan:2017osp}}
	\label{fig:BELLE_ALPS}
\end{minipage}
\end{figure}

There is an active program of searches for physics beyond the Standard Model, which include dark matter, at the LHC~\cite{Aaboud:2017phn,Aaboud:2019rtt} and there
is recently an increased interest in dedicated fixed-target experiments, with proposals being considered at CERN, JLAB, FNAL and SLAC~\cite{Akesson:2018vlm,Boyce:2012ym,Battaglieri:2017qen,Banerjee:2016tad}. 
These searches are tricky since we are trying to detect a particle that does practically not interact with matter. Collider experiments tag the production of potential
dark matter candidates by missing energy plus initial-state radiation, missing energy plus ``recoil'' visible particles or by looking for resonances at the mass of the new particle. Still, the parameter space is large, with the possibility of vector, axial-vector or scalar type of interactions, with their respective couplings as free parameters. The dark matter mass and the mass of the mediator of the interaction are further free parameters. Assumptions on some of these parameters are unavoidable, so the results presented
by accelerator experiments need to be always understood under the assumptions taken. This is important when comparing results with Astroparticle experiments, for example using limits on the $\sigma^{\mathrm{SD}}_{\chi\mathrm{-}p}$ or $\sigma^{\mathrm{SI}}_{\chi\mathrm{-}p}$ cross sections. Collider experiments do not have access directly to these quantities, but they are usually derived under some model assumptions, while Astroparticle experiments measure fluxes of particles that can be cast in a more direct way in terms of those quantities. There are different model dependencies and systematics in both approaches and comparisons must be made with care. Figure~\ref{fig:ATLAS_DM} shows the limit on the spin-independent dark matter  nucleon cross section obtained from an analysis of ATLAS data, compared with results from direct detection experiments. Collider searches can be competitive at very low dark matter masses, but the figure also shows that model dependencies can be strong (compare the blue and red lines, which correspond to
two different assumptions on the nature of the dark matter).

Electron-positron colliders provide in general a cleaner interaction environment, and this is also true for dark matter searches. Belle has developed a program of
searches for Axion-like particles (ALPs)~\cite{Dolan:2017osp} , where an ALP produced in $e^+e^-$ annihilations decay into a detectable photon pair. Figure~\ref{fig:BELLE_ALPS}
shows that Belle has complementary sensitivity to other search techniques for ALPs.

\subsection{Outlook}
The search for dark matter is a complex, multidisciplinary, experimentally driven field, since many possibilities on its nature remain
open. We have seen in the last couple of decades an impressive technological development in direct detection techniques which have increased
the sensitivity of the experiments to the dark matter cross section by several orders of magnitude. After a necessary exploratory era, 
direct detection experiments are converging to a few well established techniques, and collaborations are joining efforts for the next generation
of large volume detectors. The challenge in  direct detection is lowering thresholds and coping with the so-called neutrino floor.
Direction-sensitive detectors will be able to deal with the background of nuclear recoils induced by elastic coherent solar neutrino scattering, and
provide sensitivity to probe the very low dark matter mass region. Rather than being a hard limit, the neutrino floor will rather become a 
swampland, difficult to navigate due to the additional background, but possible with the right equipment.

Indirect dark matter searches with photons, neutrinos or cosmic rays from annihilation or decay of dark matter in the cosmos provide a complementary approach,
subject to different systematics, and are sensitive to different variables. In some cases indirect searches are competitive since they can probe different
mass regions or different dark matter properties. Yet another complementary search technique for dark  matter is performed in the controlled environment of
particle physics labs. They do not have the ability to determine the lifetime of a potential signal, but benefit from the potential
to precisely measure particle couplings and masses. 

In the big picture the question remains: is the dark sector really so simple as one stable particle, while the visible sector comprises several 
fundamental particles and families? We probably need to start considering more complex scenarios, keeping complimentary search techniques open. 
Theory can lead the way here, and there are already models assuming more complex scenarios than the standard one-stable-particle-freeze-out-from-thermal-equilibrium mechanism.

\end{document}